\documentclass[useAMS,usenatbib]{mn2e}
\usepackage{graphicx,amssymb}
\usepackage{subfigure,appendix,enumerate}
\usepackage{url,hyperref,xcolor}
\usepackage{multirow}
\usepackage{lscape}
\usepackage{natbib}

\newcommand\qbh{GS 2023+338}

\definecolor{dark-green}{rgb}{0.1,0.49,0.4}
\hypersetup{colorlinks=true, urlcolor=blue, citecolor=dark-green}

\usepackage{color}

\begin{document}
\title[The Quiescent X-ray Spectrum of Accreting BHs]{The Quiescent X-ray Spectrum of
  Accreting Black Holes} 

\author[M.T.~Reynolds et al.]
{Mark T. Reynolds$^1$\thanks{Email: markrey@umich.edu},
  Rubens C. Reis$^1$, Jon M. Miller$^1$, Edward M. Cackett$^2$ 
\newauthor Nathalie Degenaar$^1$\\ 
$^1$Department of Astronomy, University of Michigan, 500 Church Street, Ann Arbor, MI 48109\\ 
$^2$Department of Physics \& Astronomy, Wayne State University, 666 W. Hancock St., Detroit, MI 48201}

\maketitle
\begin{abstract}
The quiescent state is the dominant accretion mode for black holes on all mass scales. Our
knowledge of the X-ray spectrum is limited due to the characteristic low luminosity in
this state. Herein, we present an analysis of the sample of dynamically-confirmed
stellar-mass black holes observed in quiescence in the \textit{Chandra/XMM-Newton/Suzaku}
era resulting in a sample of 8 black holes with $\sim$ 570 ks of observations. In contrast
to the majority of AGN where observations are limited by contamination from diffuse gas,
the stellar-mass systems allow for a clean study of the X-ray spectrum resulting from the
accretion flow alone. The data are characterized using simple models. We find a model
consisting of a power-law or thermal bremsstrahlung to both provide excellent descriptions
of the data, where we measure $\rm \Gamma = 2.06 \pm 0.03$ and $\rm kT =
5.03^{+0.33}_{-0.31}~keV$ respectively in the 0.3 -- 10 keV bandpass, at a median
luminosity of $\rm L_x \sim 5.5\times10^{-7}~L_{Edd}$. This result in discussed in the
context of our understanding of the accretion flow onto stellar and supermassive black
holes at low luminosities.\\
\end{abstract}
 

\section{Introduction}
\begin{table*}
\label{obs_table}
\begin{center}
\begin{tabular}{lccccc}
\hline
System & Observatory & Exposure time & Net counts & Obsid & Reference \\ [0.5ex]
 & & [ ks ] & [ ct ] & &  \\ [0.5ex]
\hline\hline
GS 2023-338   & \textit{Chandra}    & 10.3 & 1693.6 & \#97   & \citet{garcia01}\\   [0.5ex]
 -- --        & \textit{Chandra}    & 55.6 & 1930.1 & \#3808 & \citet{kong02}\\   [0.5ex]
 -- --        & \textit{XMM-Newton} (pn, m1, m2) & 30.2, 34.0, 34.0 & 2032, 766, 816 & \#0304000201
 & \citet{bradley07}\\   [0.5ex]
 -- --        & \textit{Suzaku} (x0, x1, x3) & 42.3, 42.3, 42.4 & 1601.9, 1841.6,
 1941.0 & \#404059010 & this work\\   [0.5ex]
1A 0620-00    & \textit{Chandra}    & 42.1 & 120.9 & \#95 & \citet{garcia01}\\   [0.5ex]
 -- --        & \textit{Chandra}    & 39.6 & 301.7 & \#5479 & \citet{gallo06}\\   [0.5ex]
GRO J1655-40  & \textit{Chandra}    & 42.6 & 49.6  & \#99 & \citet{garcia01}\\   [0.5ex]
 -- --        & \textit{Chandra}    & 18.2 & 147.9 & \#10907 & \citet{calvelo10}\\   [0.5ex]
 -- --        & \textit{XMM-Newton} (pn) & 15.6 & 77.4 & \#011246201 & \citet{hameury03}\\   [0.5ex]
GS 1354-64    & \textit{Chandra}    & 39.5 & 273.1 & \#12741 & \citet{reynolds11}\\   [0.5ex]
 -- --        & \textit{Chandra}    & 19.8 & 136.1 & \#13720 & this work\\   [0.5ex]
 -- --        & \textit{Chandra}    & 19.8 & 112.6 & \#15576 & this work\\   [0.5ex]
XTE J1118+480 & \textit{Chandra}    & 45.8 & 69.4  & \#3422 & \citet{mcclintock03}\\   [0.5ex]
XTE J1550-564 & \textit{Chandra}    & 18.0 & 40.6  & \#3672 & \citet{corbel06}\\   [0.5ex]
 -- --        & \textit{Chandra}    & 47.8 & 133.3 & \#5190 & \citet{corbel06}\\   [0.5ex]
 -- --        & \textit{XMM-Newton} (pn) & 10.8 & 67.1  & \#0400890101 & \citet{pszota08}\\   [0.5ex]
GS 1124-683   & \textit{XMM-Newton} (pn) & 26.9 & 66.4  & \#00085960101 & \citet{sutaria02}\\   [0.5ex]
GRS 1009-45   & \textit{BeppoSax}   & 40.6 & 43.0  & \#20607001 & \citet{campana01}\\   [0.5ex]
\hline
\end{tabular}
\end{center}
\caption{The sample of quiescent BH X-ray spectra considered herein representing an
  accumulated exposure of $\sim$ 570 ks across 8 stellar mass black holes. The net
  counts are calculated after subtraction of an appropriately scaled background from a
  neighboring source free region. The \textit{Suzaku} observation of \qbh~is presented
  here for the first time. We also present 2 new \textit{Chandra} observations of GS
  1354-64. In addition, re-analysis of the \textit{BeppoSax} observation of GRS 1009-45
  reveals a significant detection in quiescence for the first time (see appendix for
  details). There are a number of quiescent systems detected by \textit{Chandra} that we
  do not use due to the low number of detected counts, e.g., GRO J0422+32 (\#676,
  \citealt{garcia01}), GS 2000+25 (\#96, \citealt{garcia01}), 1H 1705-250 (\#11041,
  \citealt{yang12}), SAX J1819.3-2525 (\#3800, \citealt{tomsick03}), XTE J1118+480
  (\#2751, \citealt{mcclintock03}), XTE J1859+226 (\# 3801, \citealt{tomsick03}) and GRO
  J1650-500 (\#7512, \citealt{gallo08}). In each case less than 20 net source counts
  were detected.}
\end{table*}

The nearby population of active galactic nuclei (AGN) is dominated by systems accreting at
low luminosity, i.e., $\rm L_x \lesssim 10^{-6}~L_{Edd}$
\citep{soria06,ho09,gallo10,pellegrini10}\footnote{Where $\rm L_{Edd} = \eta \dot{M}c^2
  \approx 1\times 10^{39} (M_{BH}/10~M_{\sun})~erg~s^{-1}$ and the accretion efficiency
  $\eta$ assumes the fiducial value of 0.1.}.  Understanding the nature of the accretion
flow in this low luminosity mode is important because the feedback generated by the black
holes operating in this kinetic mode may affect the host galaxy.  The energy output, when
integrated over the time lifetime of the black hole, may significantly affect the local
environment (e.g., \citealt{croton06}).
 
In such phases, accretion onto black holes is broadly expected to proceed via a
geometrically--thick accretion flow, but the details remain uncertain. This is partially
because observations are complicated by the contaminating components typical in the host
galaxy.  In principle, a low luminosity can be achieved by transferring very little gas
onto the black hole, or by transferring more gas in a manner that is radiatively
inefficient.  In practice, these mechanisms likely act in concert.  A modest inward flow
of gas onto the black hole may trap radiation so that the energy is advected across the
event horizon.  Numerous flavors of advection--dominated accretion flow (ADAF) models
have been developed around this idea (e.g.,
\citealt{ichimaru77,narayan94,narayan08,narayan00,igumenshschev02,proga03,blandford99}).
Alternatively, the inflow may channel most of the accretion energy into an outflow,
potentially producing relativistic jets (e.g. \citealt{markoff01}).  Models consisting
of both an ADAF-like and/or a jet component have been proposed, e.g.,
\citet{yuan09,eracleous10,nemmen14}, and are increasingly common.

Even as such models develop and progress, however, some fundamental uncertainties remain.
\citet{blandford99} noted that hot, radiation-trapping accretion flows with low mass flux
-- ADAFs -- may not be gravitationally bound to the black hole.  Thus, it is uncertain if
gas truly accretes onto the black hole itself, when the rate of mass transfer from large
radii is too meager. To this end \textit{Chandra} has recently invested over 4 Ms in an
effort to shed light on the nature of the quiescent accretion flow. Observations of the
$\rm \sim 10^6~M_{\sun}$ Galactic center supermassive black hole (SMBH) Sgr A$^{*}$
\citep{nowak12,wang13} and the $\rm \sim 10^9~M_{\sun}$ SMBH in NGC 3115
\citep{wong11,wong14} suggest that an ADAF with a significant outflowing component is
favored. These observations are supported by recent numerical work
\citep{yuan12a,yuan12b}. See the review by \citet{yuan14} for a detailed discussion of the
current status these so-called hot accretion flows.

Comparisons between, for example, Seyfert AGN and stellar-mass black holes in bright,
active states have helped to understand both source classes (e.g.,
\citealt{merloni03,falcke04}).  The same is potentially true in even greater measure for
black holes that accrete at very low rates.  Observations of Galactic black hole
binaries in the ``quiescent" state have facilitated the best constraints on black hole
accretion at $\rm L_x \lesssim 10^{-6}~L_{Edd}$ in the 0.5 -- 10.0 keV band.  The
spectrum is consistent with a power law of spectral index $\Gamma \sim 2$, though with
large uncertainties (\citealt{garcia01,kong02,reynolds11}). By itself, this does not
necessarily distinguish between different accretion models. There is some evidence that
jets may at least be an important addition to ADAF flows, based on the ability of jet
models to fit the IR portion of broad-band spectra in terms of synchrotron emission
(e.g. \citealt{markoff01,fender03,markoff05,maitra09}). However, the infrared spectrum
of Galactic stellar-mass black holes can be readily measured, whereas galactic starlight
obscures the IR spectrum of massive black holes in nearby low-luminosity AGN.

Infrared synchrotron emission in jets does not reveal the details of the regime closest
to the black hole, however, nor the mechanisms that may operate there, nor whether the
gas is bound.  Competing models predict that the apparent X-ray power law below $\sim
10$ keV is associated with either synchrotron and/or synchrotron self-Comptonization
(e.g. in a jet base; \citealt{markoff01}); or Comptonization in a hot (e.g. $\sim100$
keV) corona, possibly in concert with Bremsstrahlung emission \citep{esin97,narayan08}.
Basic models have been constructed to predict the emission line spectrum of hot, unbound
flows (e.g. \citealt{narayan99,perna00,xu11}); but reported limits are not very
constraining \citep{bradley07}.

In short, tremendous gaps remain in our understanding of accretion onto black holes at
low rates of mass transfer, and observations of stellar-mass black holes may be the most
practical way forward.  With the goal of understanding what the best current X-ray data
are able to reveal, we have systematically analyzed all available spectra of the
dynamically confirmed stellar-mass black holes in quiescence.  We report continuum
fitting results and limits on emission lines.  Leveraging these results, we also lay out
the kind of observations that are necessary to make progress using current X-ray
observatories.

In this paper, we describe an analysis of all archival observations of quiescent stellar
mass black holes observed to date.  The analysis presented herein provides the best
currently available constraints in the quiescent accretion spectral shape. In \S2, we
describe the observations and spectral extraction. We proceed to analyze the data in
\S3. In \S4, these results are discussed.

\begin{figure}
\begin{center}
\includegraphics[height=0.36\textheight,angle=-90]{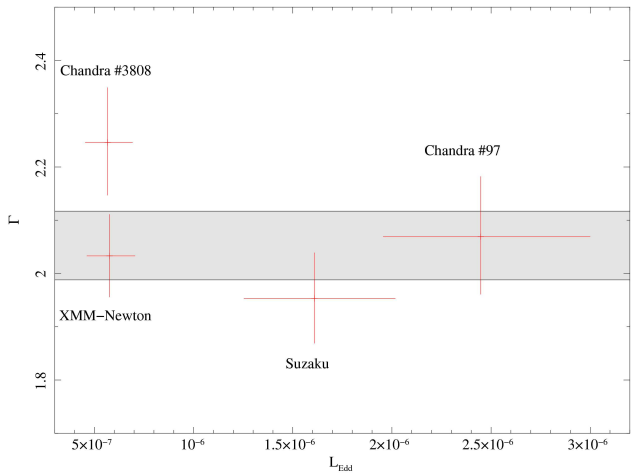}
\caption{Eddington scaled luminosity versus power-law index for GS 2023+338 assuming a
  distance and mass of 2.39$\pm$0.14 kpc \& $\rm 9.0^{+0.2}_{-0.6}~M_{\sun}$
  \citep{millerjones09,khargharia10}. The column density is assumed to remain constant
  between each observation. The gray band denotes the result of a joint fit all 4
  observations simultaneously, which supports a power law index $\Gamma =
  2.05^{+0.7}_{-0.6}$ (90\% confidence level).}
\label{gs2023_ledd_gamma}
\end{center}
\end{figure}

\section{Observations}
In this study, we consider all observations of dynamically confirmed Galactic stellar
mass black holes in the quiescence state ($\rm L_x \lesssim 10^{-6}~L_{Edd}$), observed
prior to 2013. Such observations have only become readily possible with the advent of
modern CCD based detectors (e.g., \textit{Chandra}/ACIS,
\textit{XMM-Newton}/EPIC-pn). The sample consists of 18 observations of 8 quiescent
black holes. In two of these observations, the system of interest was detected by
multiple detectors (\textit{XMM-Newton}/EPIC-pn/mos1/mos2 \&
\textit{Suzaku}/xis0/xis1/xis3), resulting in a final sample of 22 spectra. These are
the observations considered in \citet{reynolds11} plus the observation of 1H 1707-250
\citep{yang12}, two new \textit{Chandra} observations of GS 1354-64 and new
\textit{Suzaku} observation of \qbh~(aka V404 Cyg) both of which are presented herein
for the first time.

In Table \ref{obs_table}, we list the relevant properties of each observation, including
the final net number of counts detected from each source. The majority of the sample
have been previously published, though all data is re-reduced using the latest
calibration files in the current analysis. In all cases, our re-analysis is consistent
with the previously published results (see appendix A\ref{appendix_dr} for details). An
inspection of Table \ref{obs_table} reveals our sources to be faint, with the exception
of \qbh, which as the most luminous source clearly dominates the sample.  We begin our
analysis by considering this source in some detail. All spectral analysis is carried out
within \textsc{xspec
  12.7.0u}\footnote{\url{http://heasarc.gsfc.nasa.gov/xanadu/xspec/}}. The latest
versions (as of January 2013) of the relevant \textsc{caldb} files were also used.
Chi-squared statistics were used at all times with the data being appropriately
rebinned, see the appendix for details.

\begin{figure}
\begin{center}
\includegraphics[height=0.34\textheight,angle=-90]{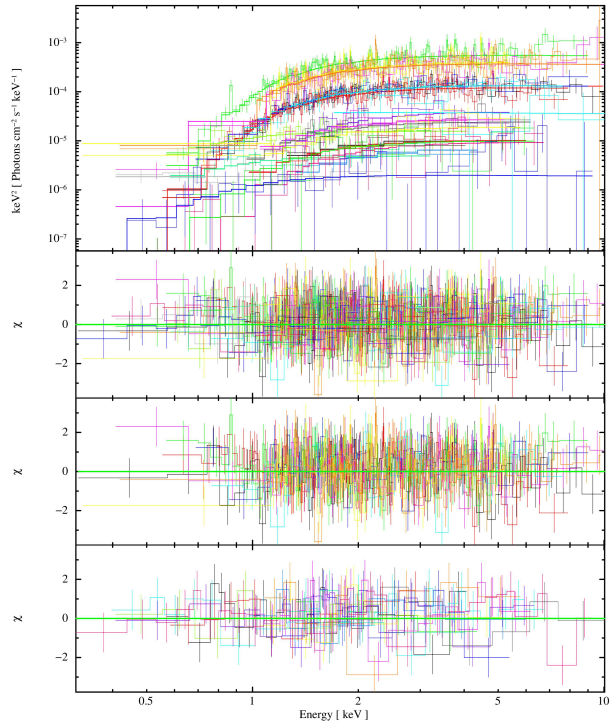}
\caption{Power law fit to all 22 spectra for the 8 systems in our sample. The lower
  panels display the residuals w.r.t. the best fit power law model: \textbf{Bottom}: Fit
  to all systems excluding GS 2023+338, where the best fit power law index $\Gamma =
  2.09^{+0.22}_{-0.21}$. \textbf{Middle}: Fit to all observations of GS 2023+338 alone,
  where the best fit power law index $\Gamma = 2.05 \pm 0.06$. \textbf{Top}: Residuals
  w.r.t. a fit to all spectra where the best fit power law index $\Gamma= 2.06 \pm
  0.06$. All errors are at the 90\% confidence level.}
\label{zall_po}
\end{center}
\end{figure}

\section{Analysis \& Results}
Previous observations of a number of quiescent black holes by \textit{Chandra} \&
\textit{XMM-Newton} have shown the 0.5 -- 10 keV spectrum to be consistent with a power
law ($\rm \Gamma \sim 2$, \citealt{garcia01,kong02,bradley07,reynolds11}). These
analysis necessarily focused on single or small samples of observations. Here we analyze
all of the spectra accumulated since the launch of \textit{Chandra} \&
\textit{XMM-Newton}.

As the sensitivity of our spectra is modest, we restrict ourselves to simple models. Two
baseline models are considered in detail: a simple power law (\texttt{pha*po}) and thermal
bremsstrahlung (\texttt{pha*brem}). The absorption by intervening neutral hydrogen is
modeled with \texttt{phabs}, where the abundances and cross-sections assumed are
\texttt{bcmc} \citep{bcmc} and \texttt{angr} \citep{aspl} respectively.

To begin, we analyze the most luminous and
hence most frequently observed quiescent stellar mass black hole \qbh.

\begin{table*}
\label{spec_params}
\begin{center}
\begin{tabular}{lcccccc}
\hline
Model & Observation & $\rm N_H$ & $\Gamma$, kT & $\rm f_x (0.3-10.0~keV)$ & $\rm \chi^2/\nu$\\ [0.5ex]  
& & $\rm [~10^{22}~cm^{-2}~]$ & \hspace{3mm}, [ keV ] & $\rm [~10^{-12}~erg~s^{-1}~cm{-2}~]$ & \\ [0.5ex]
\hline\hline
\texttt{pha*po} & \qbh\_xmm & 1.17$^{+0.07}_{-0.06}$  &  2.03$^{+0.08}_{-0.08}$  &  0.80$^{+0.03}_{-0.03}$  & 873/887 \\ [0.5ex]
 & \qbh\_cxo\#3808 &          1.17$^{+0.07}_{-0.06}$  &  2.25$^{+0.10}_{-0.10}$  &  0.79$^{+0.03}_{-0.03}$  &  \\ [0.5ex]
 & \qbh\_cxo\#97 &            1.17$^{+0.07}_{-0.06}$  &  2.07$^{+0.11}_{-0.11}$  &  3.42$^{+0.14}_{-0.14}$  &  \\ [0.5ex]
 & \qbh\_suz &                1.17$^{+0.07}_{-0.06}$  &  1.95$^{+0.09}_{-0.08}$  &  2.25$^{+0.15}_{-0.15}$  &  \\ [0.5ex]
-- -- & -- -- & -- -- & -- -- & -- -- & -- -- \\ [0.5ex]
 & \qbh\_simultaneous & 1.15$^{+0.07}_{-0.06}$  &  2.05$^{+0.07}_{-0.06}$  & -- & 894/890 \\ [0.5ex]
 & other\_simultaneous & -- &  2.18$^{+0.24}_{-0.23}$  & -- &  189/196\\ [0.5ex]
 & all\_simultaneous & -- &  2.06$^{+0.06}_{-0.06}$  & -- & 1084/1087\\ [2.5ex]
\hline
\texttt{pha*brem} & \qbh\_xmm & 0.92$^{+0.05}_{-0.04}$  &  5.38$^{+0.75}_{-0.62}$  &  0.60$^{+0.02}_{-0.02}$  & 883/887 \\ [0.5ex]
 & \qbh\_cxo\#3808 &            0.92$^{+0.05}_{-0.04}$  &  3.69$^{+0.53}_{-0.44}$  &  0.52$^{+0.02}_{-0.02}$  &  \\ [0.5ex]
 & \qbh\_cxo\#97 &              0.92$^{+0.05}_{-0.04}$  &  5.20$^{+1.19}_{-0.89}$  &  2.48$^{+0.10}_{-0.10}$  &  \\ [0.5ex]
 & \qbh\_suz &                  0.92$^{+0.05}_{-0.04}$  &  6.16$^{+1.15}_{-0.89}$  &  1.72$^{+0.11}_{-0.11}$  &  \\ [0.5ex]
-- -- & -- -- & -- -- & -- -- & -- -- & -- -- \\ [0.5ex]
 & \qbh\_simultaneous & 0.91$^{+0.05}_{-0.04}$  &  5.10$^{+0.50}_{-0.44}$ & -- & 905/890 \\ [0.5ex]
 & other\_simultaneous & -- &  3.40$^{+1.33}_{-0.82}$  & -- & 194/196 \\ [0.5ex]
 & all\_simultaneous & -- &  4.98$^{+0.47}_{-0.41}$  & -- &  1102/1087 \\ [0.5ex]
\hline
\end{tabular}
\end{center}
\caption{Results of our best fit power-law and bremsstrahlung models to the data
  presented in Table 1. The first four rows for each model detail the fit parameters for
  each of the observations of \qbh, while the sixth is for the simultaneous fit to all 4
  \qbh~observations (\S\ref{gs2023}), while row 7 denotes the simultaneous fit to all
  spectra of the other 7 black holes in our sample (\S\ref{other_systems}). The final
  row for each model reports the results of the joint fit to all 8 systems
  (\S\ref{entire_sample}). Unabsorbed fluxes are measured in the 0.3 -- 10 keV band via
  the \texttt{cflux} command. All errors are quoted at the 90\% confidence level.}
\end{table*}

\subsection{\qbh~(aka V404 Cyg)}\label{gs2023}
\qbh, being both a luminous and proximal quiescent stellar mass black hole, has been
observed by all 3 of the major X-ray observatories with CCD detectors (GS 1354-64 is in
fact more luminous but lies at a much larger distance, $\rm d \sim 25~kpc$, see
\citealt{reynolds11} for further details). In order to constrain the spectral form, the
four observations were modeled both separately and together. The observations by both
\textit{XMM-Newton} \& \textit{Suzaku} provide 3 spectra each (pn/mos1/mos2 and
xis0/xis1/xis3), which we choose not to merge. Therefore we have 4 observations
providing a total of 8 spectra (see Table \ref{obs_table}).

Firstly, the four observations are modeled with a power law, where a common Hydrogen
column density is assumed (see \citealt{miller09}), but the power law index and
normalization are allowed to vary. The power law index is tied for the sub-components of
the \textit{XMM-Newton} \& \textit{Suzaku} observations. In
Fig. \ref{gs2023_ledd_gamma}, we plot the measured power law index versus the luminosity
for each observation. With the exception of a single \textit{Chandra} observation
(\#3808), the other 3 are all consistent with $\Gamma \sim 2$ at the 90\% confidence
level (plotted), and all four observations agree at the 2$\sigma$ confidence
level. Based on this, the data were refit, only now assuming both a common value for the
column density \textit{and} the power law index. We measure a best fit value for the
power law index of $\rm \Gamma = 2.05 \pm 0.06~(90\%)$, indicated by the gray filled
region in Fig. \ref{gs2023_ledd_gamma}. Repeating the above analysis with the thermal
bremsstrahlung model reveals a best fit temperature of $\rm kT = (5.1 \pm
0.5)~keV~(90\%)$. The relevant parameters for the above spectral fits are listed in
Table \ref{spec_params}, where we note that both models return statistically equivalent
fits.

As line emission is predicted from the ADAF \citep{narayan99,perna00,xu11}, we
additionally tested for the presence of an emission lines in the iron K region of the
spectrum. To do this, narrow Gaussians ($\sigma = 0$) were added to the best fit
power-law model above consistent with an origin from Fe K$\alpha$ (6.4 keV), Fe XXV (6.7
keV), and Fe XXVI (6.97 keV). We find an upper limit to the equivalent width of emission
at the position of these lines of $\leq$ 180 eV, $\leq$ 125 eV, and $\leq$ 160 eV
respectively. These results are consistent with the previous
\textit{XMM-Newton} observations of \citet{bradley07}.

\begin{figure}
\begin{center}
\includegraphics[height=0.34\textheight,angle=-90]{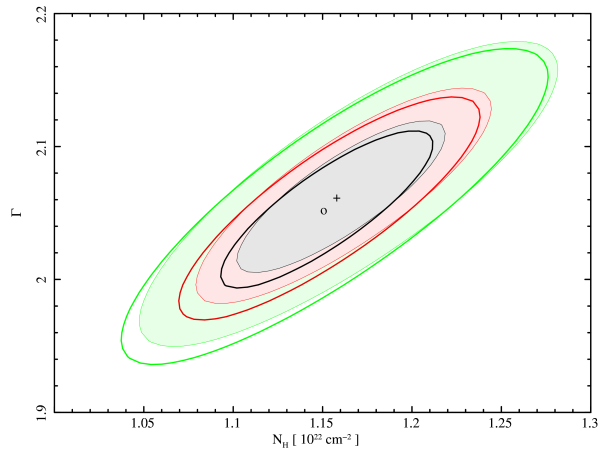}
\caption{Error contour for the power-law fits to \qbh. The thick unfilled contours
  denote the allowed parameter space when considering the data from \qbh~alone, while
  the filled contour illustrates the change in the allowed parameter space when we
  simultaneously fit all 8 systems. The black, red and green contours denote the
  1$\sigma$, 2$\sigma$ and 3$\sigma$ error contours respectively, calculated via the
  \texttt{steppar} command in \textsc{xspec}.}
\label{gs_contour_plots}
\end{center}
\end{figure}

\subsection{Other systems}\label{other_systems}
Our sample contains seven additional systems, resulting in 14 spectra in total (see
Table \ref{obs_table}). The quality of the spectra for each of these systems is
significantly less than that for \qbh~above. As such, fits to individual spectra result
in a poorly constrained spectral shape, e.g., see
\citet{garcia01,kong02,reynolds11}. Nonetheless, previous analysis point to a power law
index of $\sim 2$, as measured for \qbh~above, but with large uncertainties. Here, we
assume the quiescent spectral shape to be common to all seven quiescent black holes.

Initially, we fit the spectra with a power law model allowing the column density ($\rm
N_H$) and normalization to vary, but assuming a common power law index ($\Gamma$). The
column densities for XTE J1118+480 and GRS 1009-45 are frozen at their likely value of
${\rm N_H=10^{20}~cm^{-2}, N_H=10^{21}~cm^{-2}}$ \citep{mcclintock03,campana01} as it is not possible to
constrain the column density in these systems with the current spectra. For the other
systems $\rm N_H$ is allowed to vary and assumed to be constant for those systems with
multiple observations \citep{miller09}, while the normalization are allowed to vary
between observations. In Fig. \ref{zall_po}, we plot all 22 spectra fit with a power law
model (see \S\ref{entire_sample}). The bottom panel displays the residuals with respect
to a power law fit to the 14 spectra described above.  We measure a best fit value for
the power-law index of $\Gamma = 2.09^{+0.22}_{-0.21}$ for these 7 quiescent black
holes.

As in the case of \qbh, the above analysis is repeated with a thermal bremsstrahlung
model (\texttt{pha*brem}), which reveals a best fit temperature of $\rm kT =
4.07^{+1.71}_{-0.99}~keV~(90\%)$. The relevant parameters for these spectral fits are
listed in Table \ref{spec_params}, where we again note that both return statistically
equivalent fits.

\subsection{The entire sample}\label{entire_sample}
The results of the previous sections demonstrate that the spectral shape exhibited by
the brightest quiescent stellar mass black hole (\qbh) is statistically consistent
with that of the fainter systems. Given this fact, we repeat the above analysis, only
now we consider all spectra, i.e., 8 spectra for \qbh~and 14 spectra for the other 7
systems. 

\begin{figure*}
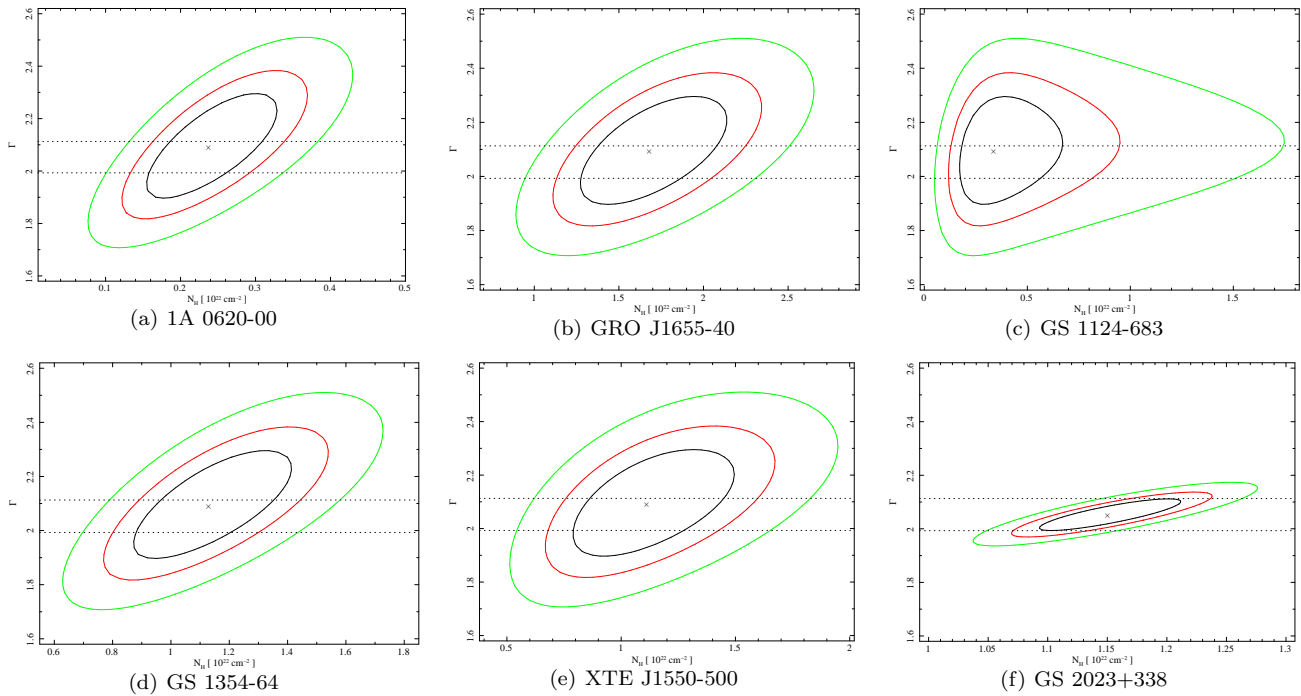

\begin{center}
\subfigure[1A 0620-00]{\includegraphics[height=0.24\textheight,angle=-90]{./fig4a.eps}}
\subfigure[GRO J1655-40]{\includegraphics[height=0.24\textheight,angle=-90]{./fig4b.eps}}
\subfigure[GS 1124-683]{\includegraphics[height=0.24\textheight,angle=-90]{./fig4c.eps}}
\subfigure[GS 1354-64]{\includegraphics[height=0.24\textheight,angle=-90]{./fig4d.eps}}
\subfigure[XTE J1550-500]{\includegraphics[height=0.24\textheight,angle=-90]{./fig4e.eps}}
\subfigure[GS 2023+338]{\includegraphics[height=0.24\textheight,angle=-90]{./fig4f.eps}}
\caption{Contour plots illustrating the allowed region in the column density power-law
  index plane for out best fit power-law model, when considering all data
  (\S\ref{entire_sample}). The \qbh~plot is for the joint fit to all 4 observations
  (\S\ref{gs2023}), while the other panels show the results of the fits to the other
  systems in the sample with measurable column density, i.e., XTE J1118+480 and GRS
  1009-45 are not plotted as the column density towards these systems is negligible. The
  contours illustrate the 1$\sigma$, 2$\sigma$, 3$\sigma$ (black, red, green) confidence
  intervals. Note the changing scales on the X-axis.}
\label{contour_plots}
\end{center}
\end{figure*}

The main panel of Fig. \ref{zall_po} plots the results of a power law fit to all 22
spectra assuming a common $\Gamma$, with the second panel displaying the residuals to
this fit. The best fit power law index is $\rm \Gamma = 2.06\pm0.06$ at the 90\%
confidence level (bremsstrahlung: $\rm kT = 5.03^{+0.47}_{-0.41}~keV$). In
Fig. \ref{gs_contour_plots}, we plot the best fit contour for $\rm N_H,~\Gamma$ for
\qbh~when it is fit alone (\S\ref{gs2023} -- thick unfilled contours) and when all 22
spectra are fit simultaneously (this section -- shaded contours). We also show in
Fig.~\ref{contour_plots} the individual best fit contour for $\rm N_H, \Gamma$ for each
source. Note the absence of XTE J1118+480 and GRS 1009-45 as the column density was
frozen for these sources.

An inspection of Fig. \ref{gs_contour_plots} \& \ref{contour_plots} reveals that the
best fit power law index, while dominated by the sensitivity of the \qbh~data is
nonetheless consistent among all 8 systems studied herein, i.e., all currently available
observations of quiescent stellar mass black holes are consistent with a relatively soft
power law ($\rm \Gamma \sim 2.06$) or a low temperature bremsstrahlung component ($\rm kT
\sim 5~keV$).

The best fit values for the column density for each system are consistent with that
previously measured at optical/IR/X-ray observations. Though in the case of GRO J1655-40
the best fit column we measure is larger than the literature value but remains consistent
within the known uncertainties, i.e., see Fig. \ref{gs_contour_plots}. At the low
luminosities characteristic of the quiescent state, we do not expect any additional
contribution from local absorbers/winds as may be present at higher luminosities. As such,
in order to obtain our final constraint on the shape of the quiescent X-ray spectrum in
accreting stellar mass black holes, the column density is fixed at the best values in
Fig. \ref{gs_contour_plots} and the spectra are re-fit. The resulting best fit power-law
index and bremsstrahlung temperature are consistent with those measured above, i.e., $\rm
\Gamma = 2.06\pm0.03$ and $\rm kT = 5.03^{+0.33}_{-0.31}~keV$. This demonstrates that the
measured spectral shapes are dominated by the intrinsic emission process and not due to
the influence of absorption at lower energies.  We note that changing the column densities
to precisely agree with the literature values does not modify these results.

In Fig. \ref{luminosity_distribution}, we plot the unabsorbed luminosity distribution in
the 0.3 -- 10 keV band for the 8 black holes in our sample, both as a pure luminosity and
an Eddington scaled quantity.  Here the Eddington scaled luminosity is the most
informative as it facilitates comparison of the observed systems. The distribution of
dynamically confirmed stellar mass black holes is observed to have a median luminosity of
$\rm L_{Edd} \sim 5.5\times10^{-7}$, though we note this is dominated by the
\qbh~observations, i.e., $\rm < L_{Edd} > \sim 2\times10^{-6}$. A clear tail to lower
luminosities is also noted, for example, 1A 0620-00 is detected at a luminosity of $\rm
\sim 10^{-8}~L_{Edd}$.  For comparison, we indicate the observed luminosity of Sgr A$^*$
in both the quiescent and flaring state\footnote{Where we have corrected the 2.0 -- 10.0
  keV luminosities from \citet{nowak12} by factors of $\sim$ 8.10, 2.18 assuming $\Gamma =
  3$ and 2 for the for the quiescent and flare luminosities respectively.}, where the
thick gray band denotes the mean to peak flare luminosity range
\citep{nowak12,degenaar13}. The flaring luminosity of Sgr A$^*$ approaches that observed
from the stellar mass sample and the observed spectral index of this flaring emission
($\rm \Gamma \sim 2$) is in agreement with the constraints on the quiescent spectral shape
presented herein. The Sgr A$^*$ quiescent spectral shape differs significantly from this
with a best fit spectral index of $\rm \Gamma \sim 3$ \citep{nowak12}.

\section{Discussion}
The overwhelming majority of black holes in the Universe accrete at a low
Eddington-scaled luminosity, yet our knowledge of the nature of the accretion flow in
this state is severely lacking. Herein, we have presented an analysis of all of the
existing data on dynamically confirmed stellar mass black holes in an effort to place
the current state of our observational knowledge in focus. Modeling the observed spectra
assuming simple power law or thermal bremsstrahlung models reveals the spectral shape
exhibited by the accretion flow at a median luminosity $\rm L_x \sim 5.5 \times
10^{-7}~L_{Edd}$ to be consistent with: $$\rm \Gamma = 2.06\pm0.03$$ $$\rm or$$ $$\rm kT
= 5.03^{+0.33}_{-0.31}~keV$$ in the 0.3 -- 10 keV bandpass, where the errors are at the
90\% confidence level. These systems provide the cleanest possible astrophysical
laboratory for the study of this low luminosity accretion flow. 

The results presented herein are consistent with previous analysis of quiescent stellar
mass BHs, e.g., \citet{garcia01,kong02,bradley07,reynolds11,plotkin13}. The quiescent
X-ray spectrum is distinct to that observed at higher luminosities, for example, in the so
called `low-hard' state a harder spectral index is observed, i.e., $\Gamma \lesssim 1.8$
\citep{mcclintock06,reynolds13}. The spectral index we measure for the stellar mass BH
sample, agrees with a scenario whereby the accretion flow changes (i.e., softens) at some
luminosity below that typically observed in the low-hard state ($\rm \lesssim
10^{-2}~L_{Edd}$). Detailed study of the decay from outburst of a number of stellar mass
BHs suggest that this softening occurs at luminosities $\rm \lesssim 10^{-5}~L_{Edd}$
\citep{corbel06,plotkin13}. Unfortunately, such soft spectra alone cannot uniquely
determine the nature of the underlying accretion flow as both ADAF, jet, and hybrid models
are capable of producing the observed power law index ($\rm \Gamma \sim 2$) in the X-ray
bandpass, e.g., \citet{esin97,markoff01b,yuan05a}.

Given the apparent scale invariant nature of the accretion flow onto black holes as
evidenced by, for example, the fundamental plane \citep{merloni03,falcke04}, it is worth
considering the similarity of the observed quiescent spectral shape of the stellar mass
black hole sample presented herein to that observed from quiescent AGN in more detail.

\subsection{Quiescent SMBHs}
The stellar mass sample studied herein lie at a median luminosity of $\rm L_x \sim 5.5
\times 10^{-7}~L_{Edd}$. The observed spectral index ($\rm \Gamma \sim 2.06$) is
consistent with that observed in large samples of nearby AGN at similar Eddington
ratios, e.g., \citet{gallo10,gultekin12}. Knowledge of the quiescent spectral shape is
important as conclusively identifying low-luminosity AGN emission is difficult, e.g.,
\citet{miller12} find a lower limit of 45\% for the occupation fraction in a sample of
Virgo cluster galaxies. The difficulty of identifying the AGN component will only
increase, as in future surveys normal galaxies will be the dominant population, in
contrast to current deep Chandra fields which are dominated by AGN, e.g.,
\citet{lehmer12}.

\begin{figure*}
\begin{center}
\subfigure{\includegraphics[height=0.27\textheight]{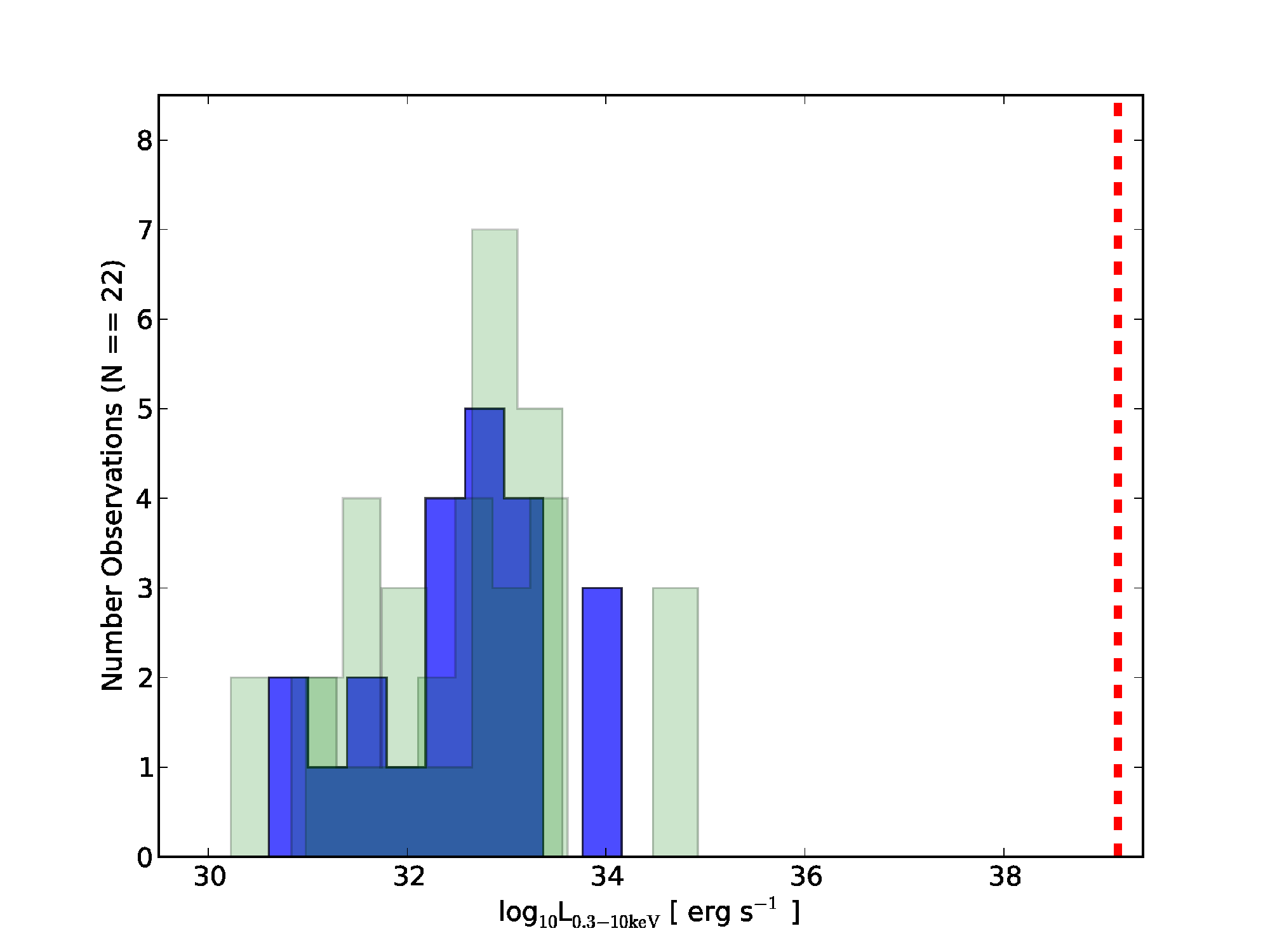}}
\subfigure{\includegraphics[height=0.27\textheight]{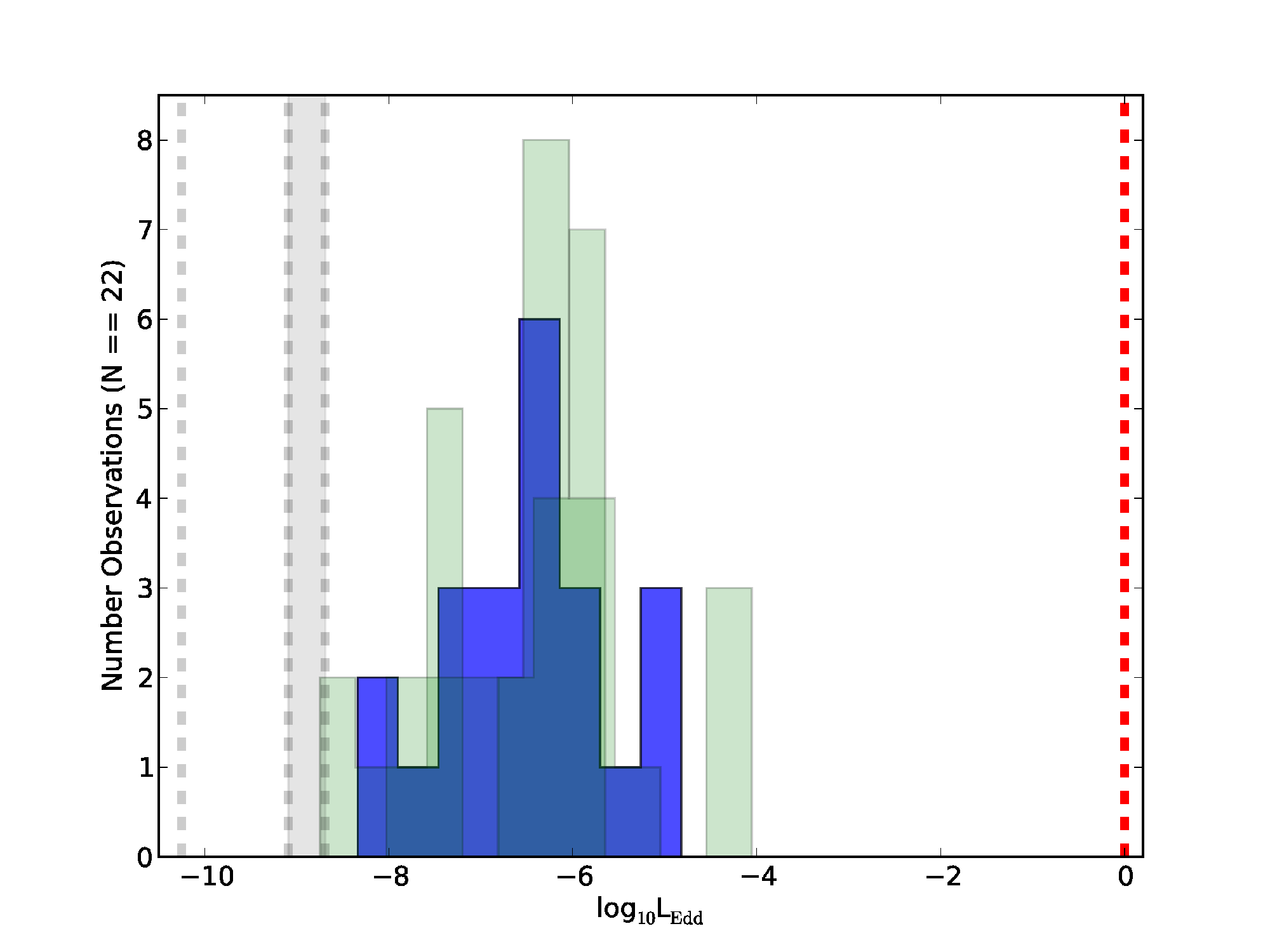}}
\caption{\textbf{Left:} Luminosity distribution for the 8 systems in our sample. The
  dashed red line represents the Eddington limit assuming a 10 $\rm M_{\sun}$ black
  hole.  \textbf{Right:} Eddington scaled luminosities corresponding to the luminosities
  in the left panel. The red dashed line represents the Eddington limit while the dashed
  gray lines represent the flaring and quiescent luminosity of Sgr A$^*$ from
  \citet{nowak12}.  The blue histogram represents the luminosity calculated assuming the
  best fit power-law model in Table \ref{spec_params}, while the green histograms
  represent the luminosities assuming the 90\% upper and lower limits.}
\label{luminosity_distribution}
\end{center}
\end{figure*}

Further constraints on the nature of the quiescent accretion flow have been provided by
observations of nearby SMBH, most notably Sgr A$^*$ \citep{nowak12,neilsen13,wang13} and
NGC 3115$^*$ \citep{wong11,wong14}. Sgr A$^*$ has recently been studied in great detail
with \textit{Chandra} where it has been found that the quiescent X-ray emission from Sgr
A$^*$ is consistent with a power law of spectral index $\Gamma \sim 3$ \citep{nowak12},
whereas the flare spectrum is consistent with a harder power-law spectral index $\Gamma
\sim 2$ \citep{nowak12,neilsen13}, consistent with the spectrum observed from the Galactic
stellar mass sample. The \textit{Chandra} campaign detected 39 discrete flaring events
from Sgr A$^*$ and although they exhibited a variation in peak luminosity by over a factor
of 10, no evidence for statistically significant evolution in the spectral shape of the
flare emission was found \citep{nowak12,neilsen13}. Similar results on the shape of the
flare spectrum have been suggested by previous observations from multiple X-ray
observatories, e.g., \textit{XMM-Newton} \citep{porquet03,porquet08} and \textit{Swift}
\citep{degenaar13}.

Based on archival CCD resolution \textit{Chandra} observations \citet{sazonov12} claim the
quiescent emission from Sgr A$^*$ to be contaminated by line of sight emission from
coronally active stars/gas implying that the true quiescent X-ray spectrum from Sgr A$^*$
was unobserved.  However, a recent analysis of the high resolution X-ray spectrum provided
by the gratings onboard \textit{Chandra} has failed to detect the predicted Fe K$\alpha$
emission line expected if emission from coronally active stars dominated the measured
X-ray spectrum, suggesting that such emission, if present, contributes no more than 25\%
of the observed spectrum \citep{wang13}. Detailed modeling of the latest \textit{Chandra}
grating spectrum favors an ADAF-type solution for the observed quiescent emission. Of
particular note, the favored solution is inconsistent with a classic ADAF where the low
luminosity is due to the advection of energy in the accretion flow (e.g.,
\citealt{narayan94}) and instead favors a scenario where a large amount of the accreted
material is ejected in an outflow (e.g., \citealt{blandford99}), hence limiting the
luminosity to be $\rm \lesssim 10^{39}~erg~s^{-1}$. The shape of the \textit{Chandra}
spectrum is dominated by bremsstrahlung emission from the inner accretion flow ($\rm
\lesssim 10^2~R_g$), but this component only contributes $\lesssim 20\%$ of the observed
X-ray luminosity within the Bondi radius (\citealt{wang13}; see also \citealt{yuan14}).

A similar deep \textit{Chandra} study of the SMBH in the galaxy NGC 3115 has been carried
out where, due to the massive $\rm 10^{9}~M_{\sun}$ black hole, the Bondi radius is
resolvable at X-ray energies \citep{wong11,wong14}. The density profile of the gas inside
the Bondi radius has been measured, where as in the case of Sgr A$^*$ an ADAF with an
outflow solution is favored. In both cases, the observed shape of the accretion flows are
consistent with that produced in updated numerical models of low-luminosity accretion
flows, which find outflows to be a ubiquitous property \citep{yuan12a,yuan12b}.

Sgr A$^*$ is by some distance the best studied quiescent supermassive black hole, and we
compare the observed quiescent accretion flow properties of this SMBH to the stellar mass
sample in more detail below.

\subsection{Quiescent stellar mass BHs and Sgr A$^*$ flares}
The flaring activity observed from Sgr A$^*$ is less luminous than the observed
distribution of stellar mass systems (see
Fig. \ref{luminosity_distribution}). Nonetheless, the observed spectral shapes are
consistent with each other ($\Gamma \sim 2$), suggestive of a common X-ray emission
mechanism.

\qbh~is the best studied quiescent stellar mass black hole at X-ray wavelengths and it is
studies of this system that have provided the best constraints on the nature of the
quiescent accretion flow. Early studies envisaged a pure ADAF model, e.g,
\citet{narayan97}; however, the presence of a radio jet in this system and numerous
quiescent SMBHs suggested a more complicated picture. Coupled with observations of Sgr
A$^*$, which constrained the accretion rate to be lower than that expected for a classical
ADAF model \citep{bower03}, this lead to the development of hybrid ADAF/jet models, e.g.,
\citep{yuan03,yuan05a,yuan05b}. 

The most recent constraints on the broadband spectral energy distribution of this stellar
mass black hole have been provided by \citet{hynes09}, where it was found that the radio
and X-ray emission are uncorrelated whereas the optical and X-ray emission exhibit a
correlation suggesting irradiation of the accretion disk by the inner ADAF-like flow (see
also \citealt{hynes02,hynes04}). The lack of a correlation between the radio and X-ray
emission argues against the X-ray emission originating in the jet, as does the measured
low UV flux, which is inconsistent with expectations from an extrapolation of the jet
component into this bandpass \citep{hynes09}. Additionally, as has been observed for X-ray
flares from Sgr A$^*$, the spectral shape of the accretion flow in \qbh~displays no
evidence for a significant deviation from a $\Gamma \sim 2$ power-law despite the
detection of significant variability \citep{bernardini14}. These points lead to
\citet{hynes09} favoring a hybrid jet plus ADAF scenario for this system.

A similar physical picture emerges from multi-wavelength observations of Sgr A$^*$ where
flares in the near infrared (nIR $\sim$ 1 -- 5 $\mu$m) and X-ray occur approximately
simultaneously, whereas as the radio flares are typically delayed suggesting a more
tenuous relationship with the process producing the X-ray/nIR emission, e.g.,
\citet{yusef08,dodds-eden09}. A model in which the X-ray and nIR flares are produced via
magnetic re-connection processes in the inner ADAF has been shown to self consistently
re-produce the observed correlated nIR/X-ray flares \citep{dodds-eden10}. An alternative
model where the X-ray emission is due to the scattering of nIR photons to X-ray energies
by hot electrons in the accretion flow has also been advocated, e.g., \citet{yusef12}.

\subsection{Stellar mass BHs and Sgr A$^*$ in quiescence}
The quiescent X-ray spectrum measured from Sgr A$^*$ is distinct from that measured during
the flaring events, and that observed from the stellar mass sample presented herein. X-ray
variability in quiescence attests to continued accretion onto the black hole in the
stellar mass case. For example, the quiescent X-ray emission from \qbh~has been observed
to be variable on all timescales probed ranging from seconds to years (e.g., see
\citealt{hynes09,bernardini14}, and Table \ref{spec_params} herein). Variability on
similar timescales is also observed from additional systems where feasible, e.g.,
\citet{kong02,reynolds11}. In contrast the quiescent X-ray emission from Sgr A$^*$ is
remarkably constant, with no appreciable variation in the non-flare emission detected
\citep{wang13}. In a detailed analysis of the flare properties observed in the
\textit{Chandra} XVP program, \citet{neilsen13} find that the flaring emission component
is unlikely to contribute more than $\sim$ 10\% of the quiescent emission. This is
suggestive of further evolution of the quiescent spectrum at the lowest luminosities/mass
accretion rates.

In the quiescent state the luminosity is highly sub-Eddington, primarily due to a reduced
accretion rate ($\rm L_x << L_{Edd}$); however, the feeding mechanism is different in the
SMBH and the stellar mass BH case. The stellar systems accrete from an accretion disk that
is replenished by material from the mass donor secondary star.  In quiescence, the
accretion disk is truncated at some distance from the black hole, likely of order $\rm
10^4~R_g$ \citep{esin97}, with the region interior to this transitioning to an ADAF of
some form. Optical studies of the quiescent outer accretion disk reveal an accretion rate
onto the outer disk of $\rm \dot{M} \sim 10^{-10}~M_{\sun}~yr^{-1}$
\citep{marsh94,mcclintock95}, whereas the Eddington accretion rate is 3 orders of
magnitude higher $\rm \dot{M} \sim 10^{-7}~M_{\sun}~yr^{-1}$ (assuming a $\rm \sim
10~M_{\sun}$ BH). Additionally, observations at UV energies have revealed the presence a
significant source of UV flux likely originating from the mass transfer stream disk impact
point, e.g., \citet{mcclintock03,froning11}.

In contrast, low luminosity SMBH such as Sgr A$^*$ accrete diffuse gas from the inner
regions of their host galaxy typically by the Bondi-Hoyle mechanism at large radii ($\rm
\sim 10^6~R_g $) transitioning to an ADAF interior to this. For Sgr A$^*$ the accretion
rate is less than that suggested by the standard Bondi mechanism ($\rm \sim
10^{-5}~M_{\sun}~yr^{-1}$), and has been constrained via Faraday rotation measurements to
be $\rm \dot{M} \lesssim 10^{-7}~M_{\sun}~yr^{-1}$ \citep{bower03}, many orders of
magnitude less than the Eddington accretion rate of $\rm \dot{M}_{Edd} \sim
10^{-2}~M_{\sun}~yr^{-1}$. The mass reservoir immediately available to the stellar mass
systems is larger than that available to the SMBH in addition to being provided via an
accretion disk vs the quasi-spherical feeding suffered by the SMBHs. This larger mass
accretion rate is likely the dominant source of the enhanced variability observed from the
stellar mass systems.

The shape of the X-ray spectrum from an ADAF depends on the detailed physics of the
accretion flow, but is expected to be dominated by emission from bremsstrahlung and
compton components \citep{narayan94,yuan14}. The bremsstrahlung emission originates from
the entire radial extent of the ADAF, but the overall X-ray spectrum is typically
dominated by the compton component. The energy at which the comptonized component
dominates is dependent upon the input energy of the seed photons provided by the lower
energy synchrotron emission, which due to the mass dependent nature of the synchrotron
peak frequency peaks at optical/radio frequencies for the stellar mass/SMBHs respectively
(e.g., $\rm \nu_{synch} \sim 10^{15} M_{BH}^{-0.5}~Hz$; \citealt{quataert99}). As a
result, the comptonized emission in the SMBH case will have a characteristic temperature
lower than that for the stellar mass systems, potentially producing a systematic
difference in the observed spectra. This can allow the bremsstrahlung component to
significantly contribute to the observed shape of the X-ray spectrum from quiescent SMBHs,
whereas the stellar mass black hole spectrum will remain dominated by the compton
component.

This will change in the presence of an outflow (i.e., \citealt{blandford99}), which
results in a reduced soft seed flux for the comptonization process. Due to the
reduction/removal of the electrons producing the highest frequency synchrotron emission
that originate from the innermost radii of the accretion flow ($\rm \sim 10~r_g$), the
comptonization component will be suppressed resulting in a bremsstrahlung dominated X-ray
spectrum. In this case, the shape of the bremsstrahlung component can provide direct
constraints on the radial mass outflow rate as has been demonstrated in the case of Sgr
A$^*$ \citep{wang13}.

Finally, it is worth noting that considerable uncertainty on the composition of the
broadband spectral energy distribution at low luminosities remains. For example, the
nature of the relationship between the ADAF and the radio jet which is observed in some
quiescent systems is uncertain, e.g., both \qbh~and M87$^*$ produce collimated radio
outflows ($\rm L_x \sim 10^{-6}~L_{Edd}$). The emission from the jet component can in
cases contribute to the X-ray bandpass \citep{yuan09,eracleous10,nemmen14}, although a
significant jet contribution in the case of \qbh~is ruled out \citep{hynes09}. Evidence
for outflowing material has been detected in the inner accretion flow at radio frequencies
from Sgr A$^*$\citep{doeleman08}; however, its relationship to a radio jet such as those
observed at higher luminosities is uncertain.

\subsection{Future prospects}
The discovery of a relationship between the mass, X-ray/Radio luminosity of accreting
black holes across the mass scale \citep{merloni03,falcke04}, suggests that the process of
accretion onto a black hole is intrinsically scale invariant. Hence, future studies of
quiescent accretion flows onto \textit{both} stellar and supermassive black holes will
play a key role in our understanding of the low luminosity accretion flow. A number of
avenues suggest that further progress is imminent, for example, the `G2' object
approaching Sgr A$^*$ has the potential to provide a once in a lifetime insight into the
accretion flow onto Sgr A$^*$ \citep{gillessen12}. Should significant material accrete
onto the black hole, we may hope to measure directly the response of the accretion flow to
a known increase in mass accretion rate. Further observations may hope to reveal the
relationship between the outflow predicted at low luminosities, the numerous radio flares,
and evidence for previous activity from Sgr A$^*$ in the Galactic center (e.g.,
\citealt{wang13,clavel13}).

Of the stellar mass systems both \qbh~and 1A 0620-00 \citep{gallo05,gallo06} have been
detected at both radio and X-ray wavelengths in quiescence. However, in neither case is it
possible to resolve the observed emission. The situation is much improved in the case of a
supermassive black hole where imaging of event horizon scale emission will become a
reality in the next decade\footnote{\url{http://www.eventhorizontelescope.org}}. Current
radio observations have probed the innermost regions of the accretion flow around a number
of SMBHs, where detailed study of Sgr A$^*$ and M87$^*$ ($\rm L_x \lesssim
10^{-6}~L_{Edd}$, \citealt{dimatteo03}) have revealed evidence for outflowing material in
the inner regions surrounding the SMBH \citep{doeleman08,doeleman12}. The currently
favored ADAF models predict outflows at such low luminosity (e.g.,
\citealt{yuan12a,yuan12b}) and such observations promise to directly constrain this
emission.

Given the abundant complicating issues when attempting to constrain the low luminosity
accretion flow onto a SMBH, the advantage of utilizing the stellar mass BH sample is
clear. Unfortunately, significant expansion of the known stellar mass sample will be
difficult, given the difficulty of dynamically constraining the mass of the black hole
for sources lying in highly crowded Galactic center fields. A new avenue would be
provided by the discovery of an isolated stellar mass black hole; however, to date no
such system is known. The \textit{eRosita} mission will survey the galactic plane region
down to luminosities of $\rm 10^{33-34}~erg~s^{-1}$ and the final survey may contain a
small number of isolated black holes, but positive identification will be difficult
\citep{merloni12}, while detection of larger numbers of isolated stellar mass BHs will
likely await the completion of SKA \citep{fender13}.

The results presented herein represent the accumulation of over 10 years ($\sim$ 570 ks)
of observations with the \textit{Chandra} \& \textit{XMM-Newton} X-ray observatories. It
is clear that to significantly advance our understanding on this issue, new large scale
observing programs will be required. When considering the stellar mass black hole sample
two immediate avenues present themselves:\\ (i) Our current constraints are unable to
determine if the X-ray spectral shape exhibits curvature or is consistent with a power law
alone. Constraining the X-ray spectral shape at energies in excess of 10 keV would clearly
be of benefit here, and with the successful launch of \textit{NuSTAR} such constraints may
be forthcoming in the near future.\\ (ii) Numerous ADAF models predict the presence of
narrow emission lines (EW $\lesssim$ 30 eV) from highly ionized iron, e.g.,
\citet{narayan99,perna00,xu11}. Imprinted in these lines will lie information on the
geometry and ionization structure of the accretion flow, e.g., \citet{wang13} recently
used the detection of highly ionized Fe XXV from Sgr A$^*$, which they successfully
reproduced with an ADAF model. Crucially, this data \textit{required} an ADAF that
generated a significant outflow (e.g., \citealt{blandford99,yuan12a,yuan12b,wang13}). The
detection of outflows at such low luminosities facilitates constraints on black hole
feedback across the entire luminosity and mass range exhibited by accreting black holes.
The current best limits on line emission from the quiescent accretion flow in a stellar
mass system are provided by the \textit{XMM-Newton} observation of \citet{bradley07};
however, at typical equivalent widths $\lesssim$ 30 eV, the predicted linewidths are not
probed. A 500 ks observation with \textit{XMM-Newton} is the only way to make progress in
this area for several years to come.

\bigskip

\noindent\textbf{Acknowledgments}\\ This research has made use of data obtained from
the High Energy Astrophysics Science Archive Research Center (HEASARC), provided by
NASA's Goddard Space Flight Center. This research has made use of data obtained from the
\textit{Suzaku} satellite, a collaborative mission between the space agencies of Japan
(JAXA) and the USA (NASA). This research made extensive use of the \textit{SIMBAD}
database, operated at CDS, Strasbourg, France and NASA's Astrophysics Data System.
We thank the anonymous referee for a constructive report which improved this paper.


\begin{appendix}

\subsection{A.1) Data Reduction}\label{appendix_dr} 

\noindent All Data reduction takes place within the
\textsc{ciao4.2}\footnote{\url{http://cxc.harvard.edu/ciao/}},
\textsc{sas11.0.0}\footnote{\url{http://xmm.esa.int/sas/}}, and \textsc{heasoft
  6.12}\footnote{\url{http://heasarc.gsfc.nasa.gov/lheasoft/}} environments. We provide
further information on the specific steps undertaken to extract spectra for each
observatory used in our analysis below.\\

\noindent{\textit{Chandra}:}\\ All \textit{Chandra} observations were reprocessed with
\texttt{chandra\_repro}. Spectra were then extracted from the resulting event lists
with \texttt{specextract} assuming a 2.5\arcsec radius extraction region and an
appropriate background region from a neighboring source free region of the
detector. All spectra were binned to have a S/N = 3 per bin and we restricted
ourselves to data in the 0.3 -- 7.0 keV energy range.\\

\noindent{\textit{XMM-Newton}:}\\
Event files for all 4 sources observed by \textit{XMM-Newton} were created using
\texttt{evselect}, where we have excluded count rates greater than 0.4, 0.35 ct
s$^{-1}$ for the EPIC-pn and MOS detectors respectively, to account for flaring. 
The resulting lightcurves were inspected to confirm the absence of remaining flares
before spectra were extracted. We extracted all events with a pattern $\leq$ 4 and 12
for the EPIC-pn and MOS assuming extraction regions of radius 20\arcsec and 30\arcsec
respectively. The resulting spectra were grouped such that there were 20 counts per
bin and all good data in the 0.3 -- 10.0 keV energy range are considered.\\

\noindent{\textit{Suzaku}:}\\ Our sample contains a single \textit{Suzaku}
observation. \qbh~was observed for $\sim$ 42 ks on 2009-11-09 (PI: Reynolds). Event
files were created in the standard manner using the relevant \textit{Suzaku}
\texttt{ftools}. Spectra were extracted assuming a 4\arcmin radius extraction region
and the background was estimated from a neighboring source free region of the
detector. The resulting spectra were grouped such that there were 20 counts per
bin. Due to the high background in the XIS1 detector only data in the 0.5 -- 5.0 keV
range was considered, whereas data in the energy range 0.5 -- 10.0 keV were considered
for the XIS0, XIS3 detectors.\\

\noindent{\textit{BeppoSax}:}\\ Our sample contains one source observed with
\textit{BeppoSax} (GRS~1009-45). We followed the standard procedures described in the
\textit{BeppoSax} data analysis
cookbook\footnote{\url{http://www.asdc.asi.it/bepposax/software/cookbook/}} in obtaining
the spectra from the raw event files. To summarize, we used \texttt{xselect} to extract
both LECS and MECS source spectra with 8\arcmin\ radius and the latest backgrounds with
a similar radius were obtained from black field
images\footnote{\url{ftp://ftp.asdc.asi.it/sax/cal/bgd/}}. Counts in excess of the
background are detected in the MECS data alone.  The latest response files were also
downloaded\footnote{\url{ftp://ftp.asdc.asi.it/sax/cal/responses/98\_11}} and the
various spectra were grouped such that there were 20 counts per bin. Spectra from MECS 2
and 3 were co-added and fit in the 1.3 -- 10.0 keV energy range.

We note that previously \citet{campana01} analyzed this dataset and claimed that GRS
1009-45 was not detected, with an upper limit to the $\rm 1.6 \times
10^{-3}~ct~s^{-1}~(3\sigma)$. However, these authors estimated the background flux level
via the counts in an annulus around the source region. In contrast, following the
recommended procedure as outlined in the data analysis cookbook, utilizing a blank sky
field, we find $\sim$ 50 net counts at a position consistent with GRS 1009-45, see Table
1 for details. We note that \citet{hameury03} previously presented the results of an
\textit{XMM-Newton} observation of this system and reported an upper limit approximately
a factor of 40 less than measured here. Given the 4 years that elapsed between these
observations (1998 vs 2002) this suggests significant variability.
\end{appendix}

\vspace{1cm}
\footnotesize{This paper was typeset using a \LaTeX\ file prepared by the 
author}


\end{document}